\pdfoutput=1
\documentclass[%
 reprint,
 amsmath,amssymb,
 aps,
]{revtex4-1}

\usepackage{graphicx}
\usepackage{dcolumn}
\usepackage{bm}
\usepackage[utf8]{inputenc}

\begin{document}

\preprint{APS/123-QED}

\title{Heterogeneous nucleation on a surface with heterogeneous surface energy}

\author{Jan Kulveit}%
 \email{jk@ks.cz}%
\author{Pavel Demo}%
\affiliation{%
 Institute of Physics of the Czech Academy of Sciences\\
 Cukrovarnická 10, Prague
}%

\date{\today}

\begin{abstract}

In the standard treatment of heterogeneous nucleation on a surface, the energy of the surface is assumed to be
homogeneous. Often its value is obtained from some macroscopic measurement. We ask the question what happens if 
we consider the surface energy to be heterogeneous. This is a straightforward generalization
and may realistically be important in a number of scenarios, e.g. when the phase forming the surface is a binary alloy, 
solid solution, in presence of self-organized or artifically created patterns on the surface. 
We examine the effects of surface heterogeneity in a few scenarios in a model system, 3D lattice Ising model.
Utilizing umbrella sampling computer simulations we find the nucleation barrier can be significantly lowered in presence
of surface heterogeneity, even if the average surface energy is kept constant. 
\end{abstract}

\pacs{Valid PACS appear here}
\maketitle


\section{\label{sec:intro}Introduction}

Nucleation is a process as ubiquitous in the nature as first-order phase transitions. In the presence of heterogeneities
such as foreign surfaces, particles or pre-existing nuclei, nucleation at such sites often
makes the dominant contribution. Study of this process, called heterogeneous nucleation, is important not only for the 
fundamental understanding but also of great practical interest in technological applications. 
In the standard treatment of the topic \cite{kashchiev2000nucleation}, the energy of the surface is taken to be
homogeneous. Often the surface energy is obtained from the macroscopic measurement, as determined e.g. by the contact 
angle of a liquid drop. Obviously, this is an idealization.  

We ask the question what happens if we consider the surface energy to be heterogeneous. This is a straightforward generalization
and may realistically be important in a number of scenarios, e.g. when the phase forming the surface is a binary alloy, 
solid solution, in presence of self-organized ordered patterns, or due to intentional modification of the surface. 
It should be noted that it is a distinct case from more commonly studied problem of heterogeneous nucleation in presence of 
perturbations of the surface geometry, e.g. surface roughness, pores \cite{page2006heterogeneous,knezevic2016pore} or morphological instabilities.
To answer this question we examine the effects of surface heterogeneity in a few scenarios in a model system. 
While our initial motivation was based on an attempt to explain differences of nucleation rates of diamond on chemically identical 
substrate in one case in the form of a nanofiber and in the other of planar layer \cite{potocky2014transformation,potocky2013diamond}, due to the complexity of such system and signifficant problems with observing nanoscale nuclei in situ,
we resorted to theoretical study and computer simulations of a much simpler system, specifically Ising model in 3-dimensional 
cubic lattice.

Our answer for ``wettable'' surface is that the heterogeneity of surface energy can affect the height of the nucleation barrier 
and strong heterogeneity can significantly lower the barrier leading to much more rapid nucleation. 

We believe this can be useful both for preparation of functional surfaces and also for understanding of some discrepancies
between predicted and observed nucleation rates in cases where surface energy homogeneity assumption cannot be satisfied.

The structure of the article is as follows: First, due to use of nucleation theory in very diverse fields and resulting variance in terminology, 
we briefly mention some concepts from the classical nucleation theory. In further section we explain the choice of the 
model used and describe applied model parameters. Finally, the obtained results are discussed and summarized.

\section{\label{sec:cnt}Classical heterogeneous nucleation}

Initially, the system is in some $\alpha$-phase, which is metastable 
with regard  to the stable  $\beta$-phase. In order to change to the $\beta$-phase, first some small 
cluster of $\beta$-phase must be formed. Small clusters constantly appear because of sufficiently massive
thermal fluctuations,
but too small clusters tend to dissolve, as the gain in free energy proportional to size of the cluster
is more than compensated by the interfacial energy of newly formed surface. Thus, there is an energy
barrier, separating two attractors - the dissolution of the cluster at small size, and its growth for large
sizes. We will label the height of the barrier $\Delta G_{c}$. Clusters at the transition state on the top
of the barrier are conventionally called ``critical clusters'' and their size ``critical size''  $n_{c}$.  

During homogeneous nucleation, the process takes place in the whole $\alpha$-phase. 
In a case of heterogeneous nucleation, the clusters are formed at heterogeneities. In this paper we
consider the case of planar heterogeneous surface (labeled as $\gamma$), with $\beta$-$\gamma$ surface 
interfacial energy lower than on $\alpha$-$\beta$ interface, creating favorable conditions for nucleation.
 
\section{\label{sec:model}Model}

\subsection{\label{sec:ising}Ising model as a testbed for nucleation}

Examining theoretical improvements and more complex cases of the nucleation theory experimentally is notoriously complicated - 
individual  nuclei are usually too small to be directly observed, particularly ``in vivo'', when the nucleation process 
is happening. Often the experimentally accessible quantity is just the total nucleation rate, sometimes further obscured 
by subsequent growth processes.  Due to the exponential dependence of the nucleation rate on parameters including 
temperature and the energy barrier, small errors in experimental parameter control lead to large differences in
observed nucleation rates. It is usually very hard to distinguish experimentally  whether some proposal is really 
an improvement to the theory, or if it just happens to push the predicted nucleation rate in the ``correct'' 
direction, compensating for some unaccounted error.

For these reasons, computer simulations proved to be extremely useful in theoretical nucleation studies. We chose
2D and 3D lattice Ising model, which has been successfully used as a testbed for nucleation theory (see also our short review\cite{2015ising}). 
One of particular advantages of Ising model is the absence adjustable parameters. Several results demonstrate that provided
correct nucleus energy term, the classical nucleation theory is in good agreement with the Ising model nucleation
simulations\cite{ryu2010validity,ryu2010numerical} in the basic case of homogeneous nucleation. This makes the model 
a good basis for exploring more complex scenarios and also suggests that results obtainable even in the simple systems
will be valid in many real-world cases.

More specifically, the model system which we use is a 3-dimensional rectangular lattice of spins. For each
lattice site $k$, there is a variable $\sigma_{k}$ taking values ${-1,1}$. Spins in two adjacent sites $j,k$ interact
with energy $J\sigma_{j}\sigma_{j}$, where $J$ is an interaction strength energy, same for all neighboring pairs.
There is also an external field $h$ interacting with each spin with energy $h\sigma_{j}$, and surface energy term
$s_{l}\sigma_{l}$, for spins neighboring surface site $l$.  Consequently, the energy of the system is described by
the hamiltonian

\begin{equation}
 H = \sum_{j,k}{J\sigma_{j}\sigma_{k}} + \sum_{j}{h\sigma_{j}}  + \sum_{l}{s_{l}\sigma_{l}} 
\end{equation} 

when the first sum is over all pairs of neighboring spins, the second one is over all the spins, and the third 
term is summation over surface spins.

We can eliminate unnecessary parameters by using the coupling constant $J$ as a unit of energy. The temperature 
is measured in units of $J/k_{B}$ where $k_{B}$ is the Boltzmann constant, and the strength of magnetic field $h$ 
also in units of $J$.

We use rectangular simulation cell with surfaces on two opposing sites, and periodic boundary conditions 
in the remaining two directions parallel to the surfaces. (See fig.)  All the non-surface spins have 6 nearest neighbors,
the surface spins have only 5 neighbors.

The Ising model exhibits the famous ferromagnetic phase transition \cite{onsager1944crystal}. In 3D the critical temperature of the model
is $T_{c}=4.51 J/k_{B}$ \cite{preis2009gpu}. We explore the system at temperature $T=0.6 T_{c}=2.71 J/k_{B}$ which is well below the transition to disordered
phase. 

The temperature is above the roughening temperature $T_{r}=0.57T_{c}$ of the spin phase interfaces \cite{weeks1973structural} / if the
system was bellow the roughening temperature, we would expect cubical nuclei with flat walls (with some noise). Above the 
roughening temperature, expected nuclei are more spherical with irregular interfaces. 

The applied magnetic field is $h= 0.57 J$ in direction opposite to the initial orientation of the spins.

\subsection{Energy landscape, sampling technique and reaction coordinates}

The evolution of the system is simulated using Monte Carlo approach with the spin-flip dynamics  (see, e.g. \cite{binder2010monte}).

In a straightforward simulation, the system would spend most of the time close to the 
two attractors - the initial state with most spins in $\alpha$ phase and the final state, with almost all spins in $\beta$
phase, minus fluctuations. However, for understanding
nucleation, these states close to the two attractors are quite uninteresting, and contrarily, the transition states ``on
the top'' of the nucleation barrier are the most important. This leads to the necessity to use some advanced statistical
sampling technique, which collects more information about the nucleation barrier.

We use the umbrella sampling approach \cite{torrie1977nonphysical,frenkel2001understanding}. The idea of the method is as follows: In Metropolis Monte Carlo, in every step of the 
simulation,  flip of one spin is attempted. If the new state is energetically favorable, it is accepted. Otherwise, a random 
number is compared to the Boltzmann probability of the flip, and if the random number is larger, the flip is still accepted, 
otherwise, the spin returns to previous state. In umbrella sampling, the standard Boltzmann probability is replaced
by a factor adding a bias potential $V(\sigma)$ to the energy of the system. The potential is chosen to keep the system in a region of the energy landscape, which would normally be undersampled. A series
of such umbrella samplings may be used to explore the whole nucleation transition, and from the results we can
recover the original energy profile. For this estimate we use Benett acceptance ratio (BAR) \cite{bennett1976efficient} as implemented in
the PyMBAR code \cite{shirts2008statistically,pymbar}.

Sometimes misunderstood or neglected in nucleation studies is the important role of the reaction coordinate. The classical theory
approach can be viewed as coarse-graining the system into one dimmensional Markov chain, where some measure of the largest cluster
size is used as the coordinate. In capillarity approximation, the number of cluster particles, its radius, surface,
and surface energy are usually taken to be tied by some simple relations (e.g. $r(n)=a.n^{1/3}$, $S(n)=b.n^{2/3}$ where $r$ is radius, $S$ surface, 
$n$ the number of monomers forming the cluster, for spherical nuclei), any of such variables may be used as the reaction coordinate. 
However, in realistic situations, and also in the simple Ising model, it is clearly not obvious that this mapping 
of a huge configuration space into one coordinate is enough. E.g. two nuclei consisting of the same number
of monomers but with widely different surface areas may be one more likely to dissolve, the other to grow. 
Fortunately, in the case of Ising model it was demonstrated the size of the cluster seems to be a good reaction 
coordinate \cite{kuipers2009non}.

An important pitfall in this mapping is a correct counting of 
clusters: often used is the geometrical cluster counting, in which case a spin is considered a member of a cluster if any of its 
neighboring spins is also a member of the cluster.
This is straightforward, fast to compute, and unfortunately inappropriate way how to project the space - geometrical clusters
are unphysical, and not reflecting correctly the thermodynamics of 
the system \cite{schmitz2013monte}. We use adjusted cluster counting, in which membership of a spin in a physical cluster is determined by following 
procedure: we consider each bond between neighboring spins with the same orientation active only with probability
\begin{equation}
 p(T) = 1 - \exp{(-2J/k_{B}T)}
\end{equation} 
where $J$ is the interaction strength energy constant, and $T$ is temperature. Then, when we discover the clusters by following bonds,  
we test if the probability $p$ is greater than a random number $a \in (0,1)$, and depending on the result extend the cluster only 
when the bond is ''active``. A geometrical cluster hence can contain more than one physical clusters, and an element of randomness is
introduced into the cluster counting procedure.

\subsection{Model of surface heterogeneity}.

\begin{figure}
 \includegraphics{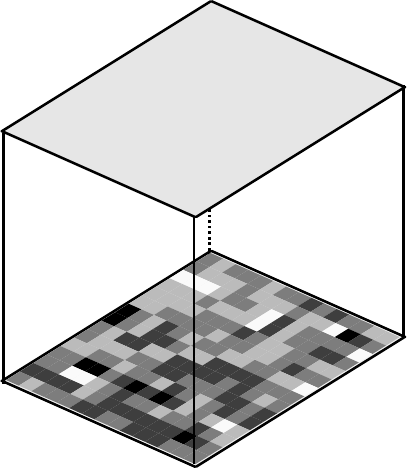}
 \caption{Schematics of the simulation cell: the lower heterogeneous surface has favorable conditions for nucleation. The
 upper surface has surface energy preventing nucleation. The boundary conditions are periodic in remaining directions. }
 \label{fig: simcell}
\end{figure}

\begin{figure}
 \includegraphics{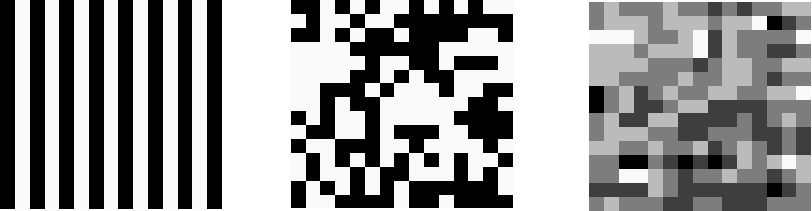}
 \caption{Three examined cases of surface heterogeneity - shade of gray represents the difference from a homogeneous 
 surface, which would be uniform gray. From the left: 1. regular stripes 2. random pattern of species 3. correlated random variable}
 \label{fig: patterns}
\end{figure}

As mentioned above, we model the surface by including a surface term $\sum_{l}{s_{l}\sigma_{l}}$ where energy 
of the bond between surface surface site $l$ and the attached spin may differ from site to site. 
The surface energy $s_{l}$ may be split into homogeneous part $s_0$ and the variable, heterogeneous part $a.p(l)$, 
where $a$ is an amplitude (or ``strength'') of the heterogeneity, and $p(n)$ is a pattern.

We examine three patterns of heterogeneity and for each pattern we run a series of simulations where
the amplitude of the heterogeneity is gradually increased. First we use regular stripes of sites with lower 
and higher surface energy ($p(n)\in\{-1,1\}$).
Second random pattern of sites with lower and higher surface energy ($p(n)\in\{-1,1\}$).
and finally, we applied surface where its energy is a random variable with binomial distribution ($p(n)\in(-1,1)$).

\section{Results}

The model was run in a cubic cell consisting of $24^{3}$ spins. One boundary of the cell is generated by the surface with
the above described heterogeneous energy. On the opposite side of the cell we created a ``non-wetting'' surface with 
field orientation reversed, strongly unfavorable for nucleation. 

In all of three surface patterns, we observed the same pattern: clusters form by heterogeneous nucleation,
and the nucleation barrier height depends on amplitude of the heterogeneities and also on a characteristic length 
scale of the pattern. Heterogeneity of the surface can cause marked decrease in the nucleation barrier.

\subsection{Regular stripes}

\begin{figure}
 \centering
 \includegraphics{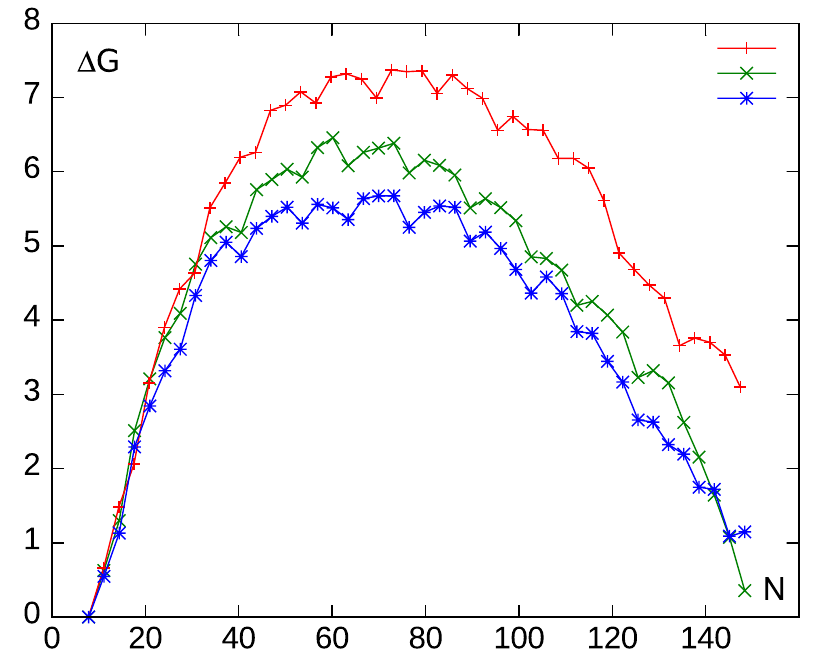}
 \caption{ \label{fig: plot-stripes}Decrease of nucleation energy barrier $\Delta G$ with increasing ``contrast'' of surface pattern in 
 case of surface patterned with regular stripes. Heterogeneity amplitudes from the top: 0.2, 0.4, 0.8 in units of $h$. Size of the cluster $N$ is simply the number 
 member spins.}
\end{figure}

In this case the inhomogeneity consists of regular stripes 2 lattice constant wide. When the 
inhomogeneity is introduced, first we observe small increase of nucleation barrier,   
but with increasing amplitude the trend is soon reversed and the nucleation barrier decreases. See Fig~\ref{fig: plot-stripes}. In the presence of
strong inhomogeneity, the barrier is almost halved. Due to the exponential dependence of nucleation rates
on barrier height, this means nucleation rate can be increased by several orders of magnitude.

\subsection{Random pattern}

\begin{figure}
 \centering
 \includegraphics{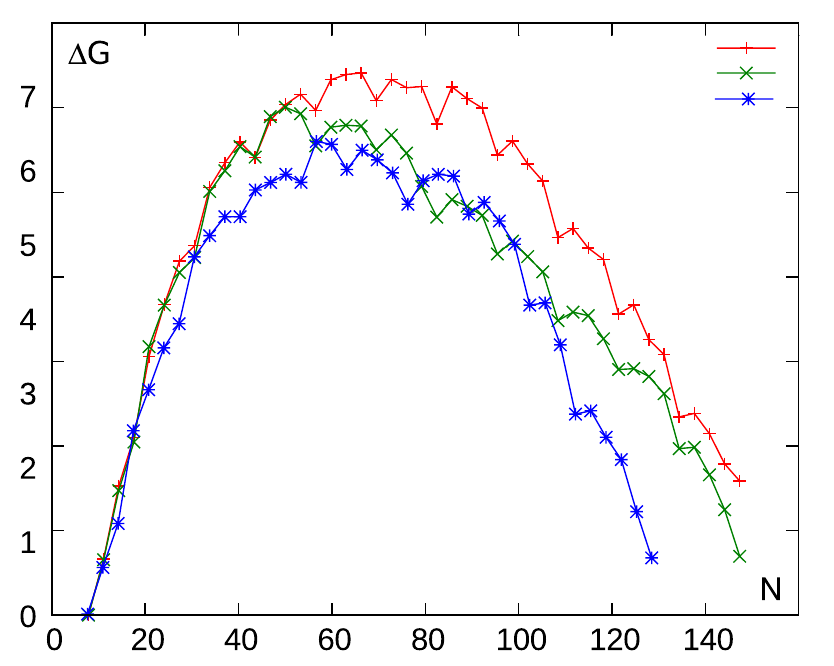}
 \caption{ \label{fig: plot-ran}Decrease of nucleation energy barrier $\Delta G$ with increasing ``contrast'' of surface pattern in 
 case of random surface. Heterogeneity amplitudes from the top: 0.2, 0.4, 0.8 in units of $h$. Size of the cluster $N$ is simply the number 
 member spins.}
\end{figure}

This heterogeneity is a random pattern of sites with two different energies. (See Fig~\ref{fig: plot-ran}). The change of the nucleation barrier 
is observeable, but less pronounced than in previous case. Again with increasing inhomogeneity the nucleation barrier is lowered.

\subsection{Random surface}

\begin{figure}
 \centering
 \includegraphics{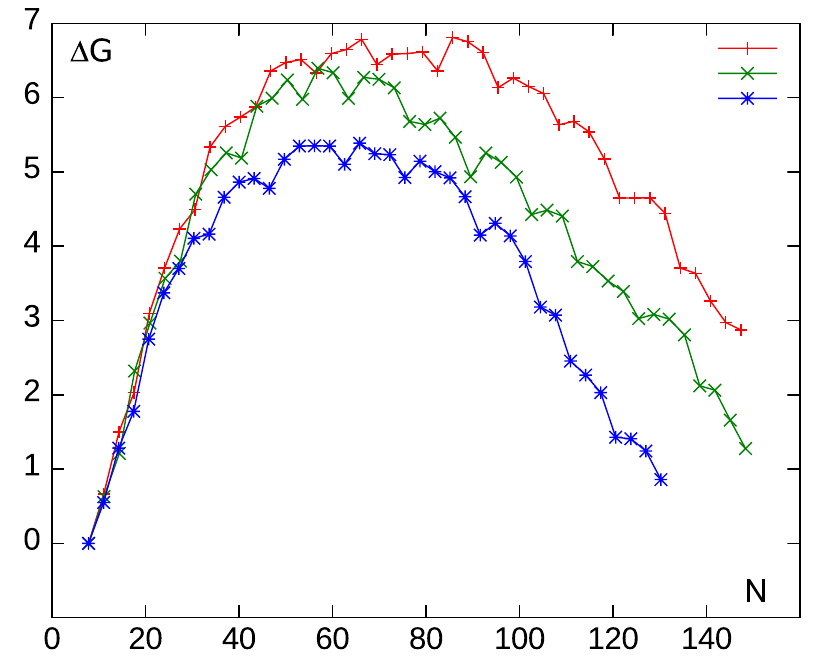}
 \caption{ \label{fig: plot-cor}Decrease of nucleation energy barrier $\Delta G$ with increasing ``contrast'' of surface pattern in 
 case of random surface. Heterogeneity amplitudes from the top: 0.2, 0.4, 0.8 in units of $h$. Size of the cluster $N$ is simply the number 
 member spins.}
\end{figure}

Here the introduced inhomogeneity is a pattern generated by addition of noise at length scales of 1,2,4 lattice units, normalized 
so the mean value of the noise across the surface is zero. See Fig~\ref{fig: patterns}. Again, with increasing 
amplitude of the inhomogeneity, we observe decrease in the nucleation barrier. See Fig~\ref{fig: plot-cor}.

\section{Concluding discussion}

In the present work we have studied heterogeneous nucleation in Ising model on simple cubic lattice, with a planar wall.
While usually such walls are considered like homogeneous, we examined several cases of surface with heterogeneous energy.

The simulation results indicate the nucleation barrier can be substantially reduced by the inhomogeneities of the surface
energy on the surface where the nucleation takes place. While our model is realively simple, we expect this conclusion is true also 
for more realistic nucleation scenarios. It shows that on surfaces with nanoscale heterogeneities it is insufficient to use
the average surface energy, obtainable from macroscopic measurement.

Intuitively, this can be understood as an ability of the nucleation process to take advantage of sites with lower surface energy,
even when the average surface energy remains stable.

From a more abstract viewpoint we can ask how a distribution of surface energies on a heterogeneous surface will influence
nucleation, and if the classical model which uses average of the energy can be improved by addition of some simple term
describing heterogeneity.

\section*{Acknowledgements}

This work has been supported by projects of GACR (The Czech Science Foundation) 15-12420S and P108/12/0891. The authors would also 
like to acknowledge the contribution of the COST Action CM1402 Crystallize.

\bibliography{heterohetero}

\end{document}